\documentclass[11pt]{amsart}

\usepackage{epsfig}
\usepackage{graphics}
\usepackage{graphicx}
\usepackage{amsmath}
\usepackage{url}
\usepackage{subfigure}

\newcommand{\W}{{\footnotesize waves}}
\newcommand{\C}{{\footnotesize oscillators}}

\pdfoutput=1

\begin{document}

\title{On oscillators in phyllosilicate excitable automata}

\author{Andrew Adamatzky}

\address{University of the West of England, Bristol, United Kingdom}
\email{andrew.adamatzky@uwe.ac.uk}

\date{\today}

\begin{abstract}

\vspace{0.5cm}

\noindent
Phyllosilicate is a sheet of silicate tetrahedra bound by basal oxygens. A  phyllosilicate excitable automaton is a regular network of finite state machines, which mimics structure of a silicate sheet. A node of the silicate sheet 
is an automaton, which takes resting, excited and refractory states, and updates its state in discrete time 
depending on a sum of excited states of its three (silicon automata) or six (oxygen automata) closest neighbours. Oscillator is a localised compact configuration of non-quiescent states which undergoes finite growth and 
modification but returns to its original state in a finite number of steps. We show that  phyllosilicate excitable 
automata exhibit spiral and target waves, and oscillating localisation dynamics. Basic types of oscillators are 
classified and characterised.

\vspace{0.5cm}

\noindent
\emph{Keywords: cellular automata, silicate, oscillator, excitation} 
\end{abstract}

\maketitle

\section{Introduction}

Phyllosilicates are parallel sheets of silicate tetrahedra, they are widely present in nature and 
typically found in clay-related minerals on the Earth surface~\cite{griffen_1993,Bergaya_2006,Bleam_1993}.
Phyllosilicates are used in development of nano-materials, nano-wires and patterned surfaces for 
nano-biological interfaces~\cite{Monnier_1993,Carrado,Suh_2009}. They are also employed, in a form 
of cation-exchanged sheet silicates, as catalysts in chemical reactions~\cite{Ballantine_1984}, e.g. 
nickel phyllosilicate catalysts~\cite{Lehmann_2012,McDonald_2009,Specht_2010}. In the paper we continue lines of enquiry into space-time dynamics of cellular automata on non-orthogonal and aperiodic lattices, including triangular tessellations and Penrose tilings~\cite{bays_2007,owens_2010,goucher_2012}. We define an automaton network, where connections between finite automata are inspired by simplified structure of silicate sheets (as lattices of 
connected tetrahedra), and investigate the dynamics of excitation on the automaton networks for various excitation rules. 

The automata models studied in the paper are  abstractions of silicon sheets. The automata models do not aim to compete with mainstream computational chemistry models~\cite{richardson_2008}.  What are advantages of the automata approach? Automata models is a fast prototyping tool for an express evaluation of a space-time dynamics of a spatially-extended active nonlinear medium for different excitation rules, and prototyping of unconventional computing devices based on the nonlinear medium. Even  abstract three-state automata are not totally pointless abstractions in imitating excitations in active media. Consider, e.g., Belousov-Zhabotinsky (BZ) medium. All types of travelling waves --- spiral and target waves in excitable mode and wave-fragments (dissipative solitons) in sub-excitable mode are sufficiently imitated in Greenberg-Hasting automata~\cite{greenberg_1978}. The automaton model of BZ medium is phenomenologically equivalent to more accurate, close to physical and chemical reality, Oregonator model~\cite{field_1974}.  By analogy, studies of phyllosilicate automata may help us to get an insight into dynamics of 
localizations, defects and excitations, in the lattices of silicon tetrahedra.  By developing automata models of excitation in tetrahedra sheets we contribute towards future developments of silicate sheet based computing circuits where quanta of information are transferred and processed by stationary and travelling localised excitations~\cite{adamatzky_rda}.

\section{Phillosilicate automata}

\begin{figure}[!tbp]
\centering
\includegraphics[width=0.5\textwidth]{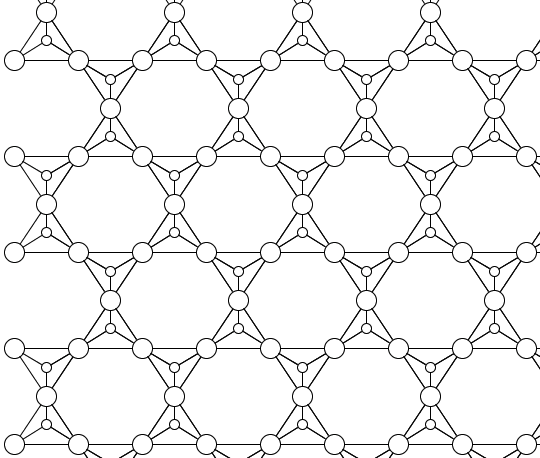}
\caption{Phillosilicate automata. Automata representing silicon are small discs, 
oxygen automata are large discs.  Unshared oxygen atoms of the tetrahedra are not shown.}
\label{latticestructure}
\end{figure}

\begin{figure}[!tbp]
\centering
\subfigure[$\sigma^t(s)>0$, $\sigma^t(o)>0$]{\includegraphics[width=0.45\textwidth]{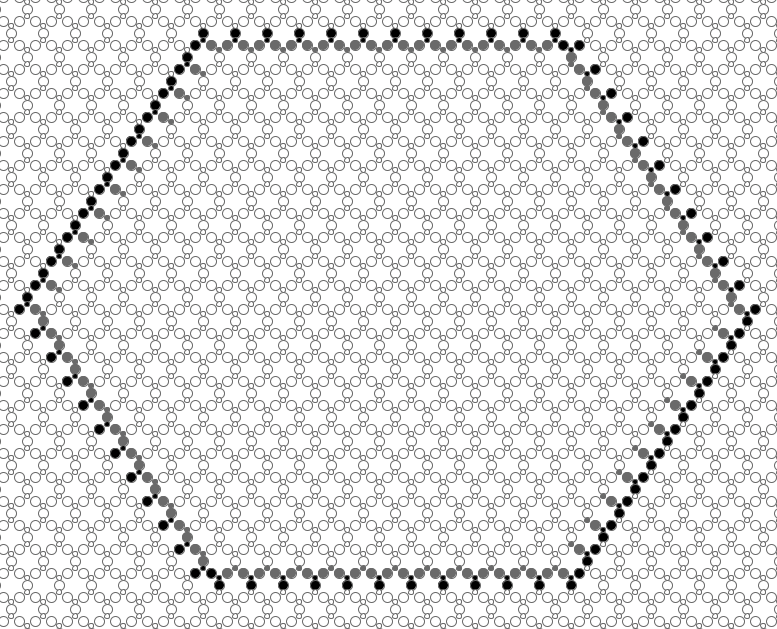}}
\subfigure[$\sigma^t(s)>0$, $\sigma^t(o)=1$]{\includegraphics[width=0.45\textwidth]{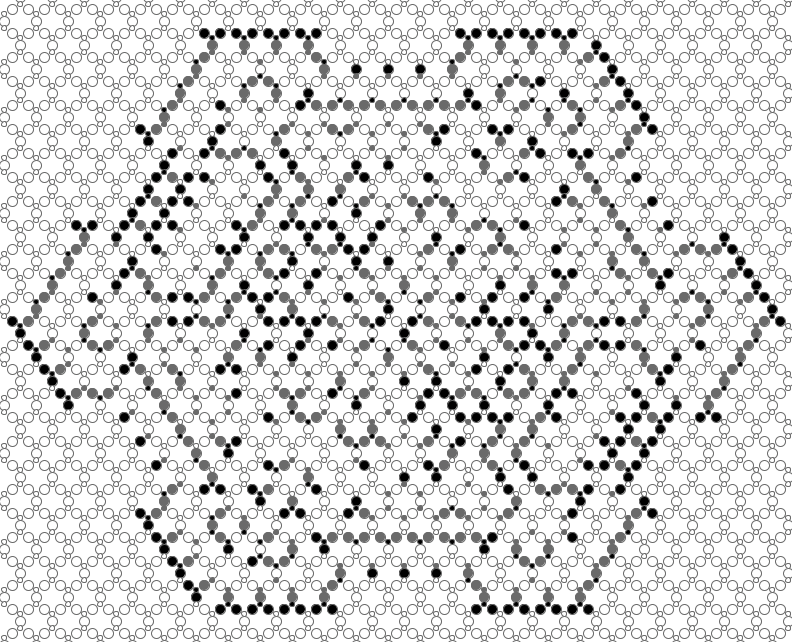}}
\subfigure[$\sigma^t(s)=1$, $\sigma^t(o)>0$]{\includegraphics[width=0.45\textwidth]{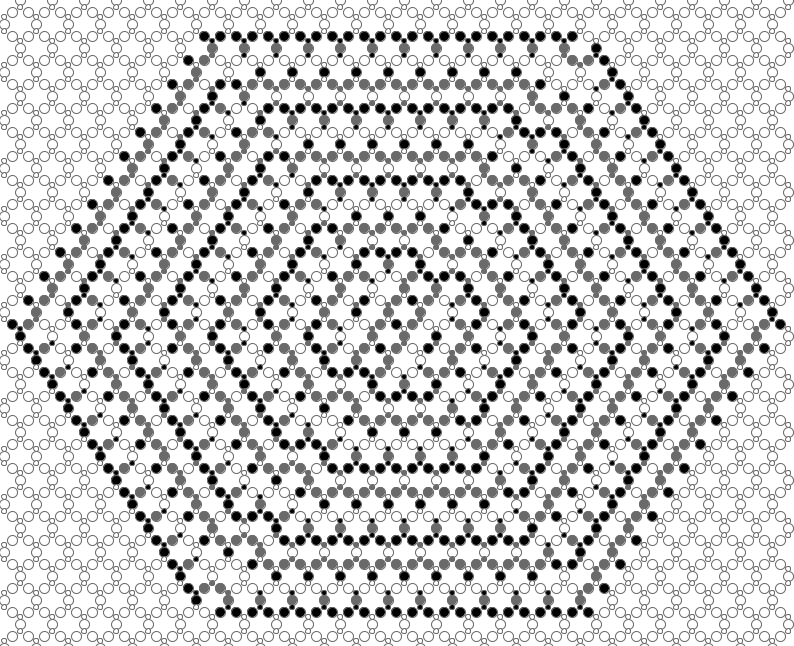}}
\subfigure[$\sigma^t(s)=1$, $\sigma^t(o)=1$]{\includegraphics[width=0.45\textwidth]{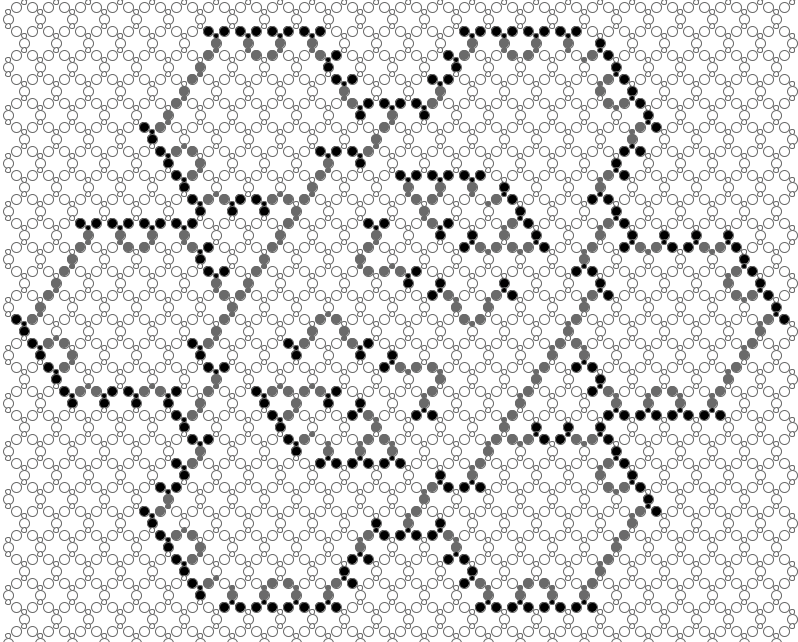}}
\subfigure[$\sigma^t(s)>1$, $\sigma^t(o)>0$]{\includegraphics[width=0.45\textwidth]{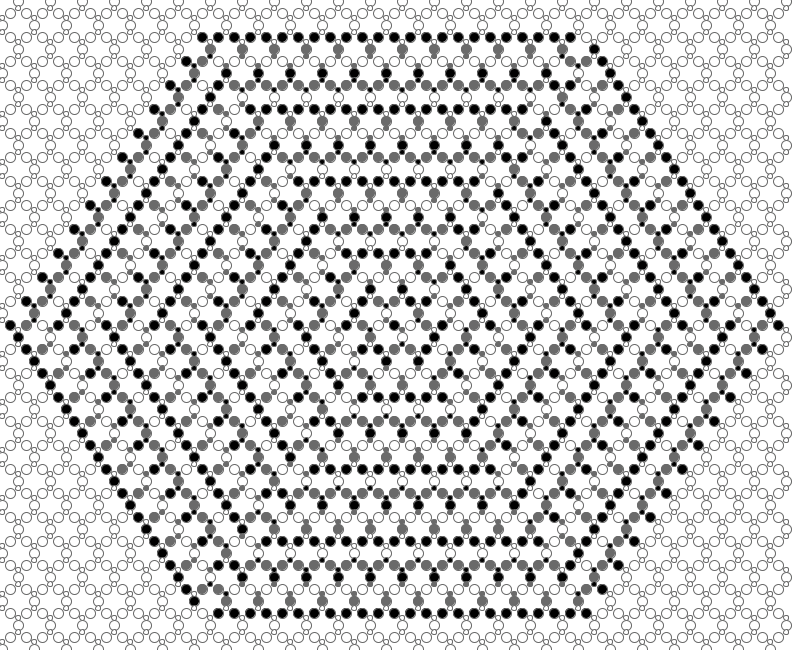}}
\subfigure[$\sigma^t(s)>1$, $\sigma^t(o)=1$]{\includegraphics[width=0.45\textwidth]{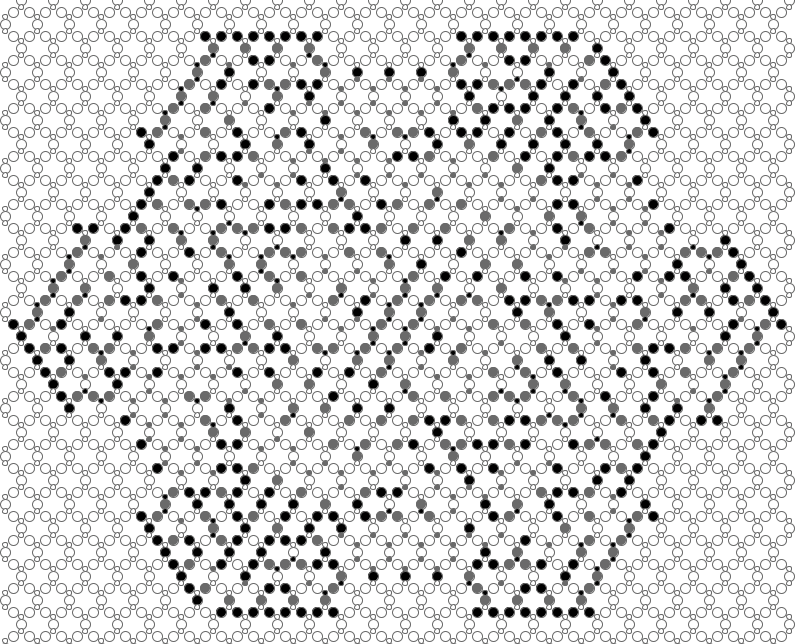}}
\caption{Configurations of excitation developed after exciting a single oxygen node in the resting lattice. 
Predicates $C_x(\sigma^t(x))$ are show in sub-figures' legends.  Excited nodes are solid discs, refractory nodes are grey discs, resting nodes are empty discs. Configurations are recorded at 25th step of the lattice iteration.  }
\label{singlestartexamples}
\end{figure}

In silicates, coordinated tetrahedra of oxygen anions are formed around each silicon cation.
In late 1920s Pauling proposed that phyllosilicates are sheets of coordinated  (SiO$_4$)$^{4-}$ tetrahedra 
units, where each tetrahedron shares its three corner (basal) oxygens of neighbouring 
tetrahedra~\cite{pauling_1930,Liebau_1985,Bleam_1993} (Fig.~\ref{latticestructure}).  Vacant oxygens of all tetrahedra point outside the lattice, away of tetrahedra; these apical oxygens are not taken into account in our present simulation. 

A phyllosilicate automaton $\mathcal A$ is a two-dimensional regular network Fig.~\ref{latticestructure} of finite state
machines. There are two types of automata in the network: silicon automata $s$ (centre vertex of each tetrahedron)  and oxygen automata  $o$ (corner vertices of each tetrahedron). A silicon automaton has three neighbours. An oxygen automaton has six neighbours (Fig.~\ref{latticestructure}).  The automata take three states   $\cdot$ (resting), $+$ (excited) and $-$ (refractory), and update their states $s^t$ and $o^t$ simultaneously and in discrete time $t$ depending on states of their neighbours. Let $\sigma^t(x)$ be a number of excited neighbours of automaton $x$, 
$0 \leq \sigma^t(s) \leq 3$ and  $0 \leq \sigma^t(o) \leq 6$. Each automaton-node $x=s, o$ of   $\mathcal A$ updates its state $x^t$ by the following rule:
\begin{equation}
x^{t+1}=
\begin{cases}
+, \text{ if } x^t=\cdot \text{ and } C_x(\sigma^t(x)) \\
-, \text{ if } x^t=+ \\
\cdot, \text{ otherwise }
\end{cases}
\label{rule}
\end{equation}
where predicate $C_x(\sigma^t(x)) \in \{(\sigma^t(x)>0), (\sigma^t(x)=1), (\sigma^t(x)>1), 
(\sigma^t(x)=2)\}$.

Exemplar configurations of excitation developed after single-node excitations are shown 
in Fig.~\ref{singlestartexamples}. A typical 'circular' wave of excitation develops when  $C(s)^t=(\sigma^t(s)>0)$ and  
$C(o)^t=(\sigma^t(o)>0)$. (Fig.~\ref{singlestartexamples}a).  Patterns of target waves are observed in situations
 $C(s)^t=(\sigma^t(s)=1)$ and  $C(o)^t=(\sigma^t(o)>0)$  (Fig.~\ref{singlestartexamples}c), 
 and  $C(s)^t=(\sigma^t(s)>1)$ and  $C(o)^t=(\sigma^t(o)>0)$  (Fig.~\ref{singlestartexamples}e). Wave-fragments are formed when $C(o)^t=(\sigma^t(o)=1)$ and  $C(s)^t \in \{ (\sigma^t(s)>0), (\sigma^t(s)=1), (\sigma^t(s)>1) \}$
  (Fig.~\ref{singlestartexamples}bdf).  
  
In scenario illustrated in Fig.~\ref{singlestartexamples}b we observe fragments of excitation waves, open ended segments of wave-fronts. These wave-fragments expand. Their ends bend backwards, split away, new wave-fragments propagate towards originally excited site. They collide with outcome wave-fragments, new wave-fragments are produced in the result of collision. Being strictly deterministic an excitation dynamics inside expanding perturbation domain might look chaotic (Fig.~\ref{singlestartexamples}b). In (Fig.~\ref{singlestartexamples}d) wave-fragments undergo a binary division at every iteration. This is because only resting nodes with exactly one excited neighbour are excited, and therefore only end of any wave-fragment transfer excitation to the resting medium; interior of the wave-fragment extinguishes.

Development of $\mathcal A$ after random (each node gets excited with probability 0.3) excitation  for all possible combinations of $C^t(x)$, $x=o, s$ is classified in Tab.~\ref{tabpatterns}. For excitation rules $\sigma^t(o) >0$ and $\sigma^t(o) =1$,  a random excitation leads to formation of generators of waves and wave-fragments. For $\sigma^t(s) >1$   and $\sigma^t(o) >1$, and  $\sigma^t(s) >2$ or $\sigma^t(o) >3$, all excitation, evoked by a random 
 stimulation, extinguishes after few iterations and the automaton returns to its resting state.

\begin{table}[!tbp]
\caption{Outcomes of random excitation of $\mathcal A$. Excitation does not persist in scenarios marked by '--'.} 
{\begin{tabular}{c|cccccc}
 &                           &                          & $\sigma^t(o)$ &                          &                          \\ \hline 
&			& $>0$ & $=1$ &$ >1$ & $=2$ & $>2$ \\ \hline
		&$>0$&   \W                       &   \W                       &  \C			 &    \C                     &         --                 \\
		&$=1$&      \W                     &  \W                         & \C 			 &  \C                        &             --             \\
 $\sigma^t(s)$ &$>1$&    \W                       & \W                          & 	 --		 &                    --      &           --               \\
&$=2$&              \W             &                   \W       & 		--	 &          --                &               --           \\
&$>2$&               \W            &              \W            & 	--		 &             --             &          --                \\
\end{tabular}}
\label{tabpatterns}
\end{table}

\section{Oscillators}

\begin{figure}[!tbp]
\centering
\subfigure[]{\includegraphics[width=0.8\textwidth]{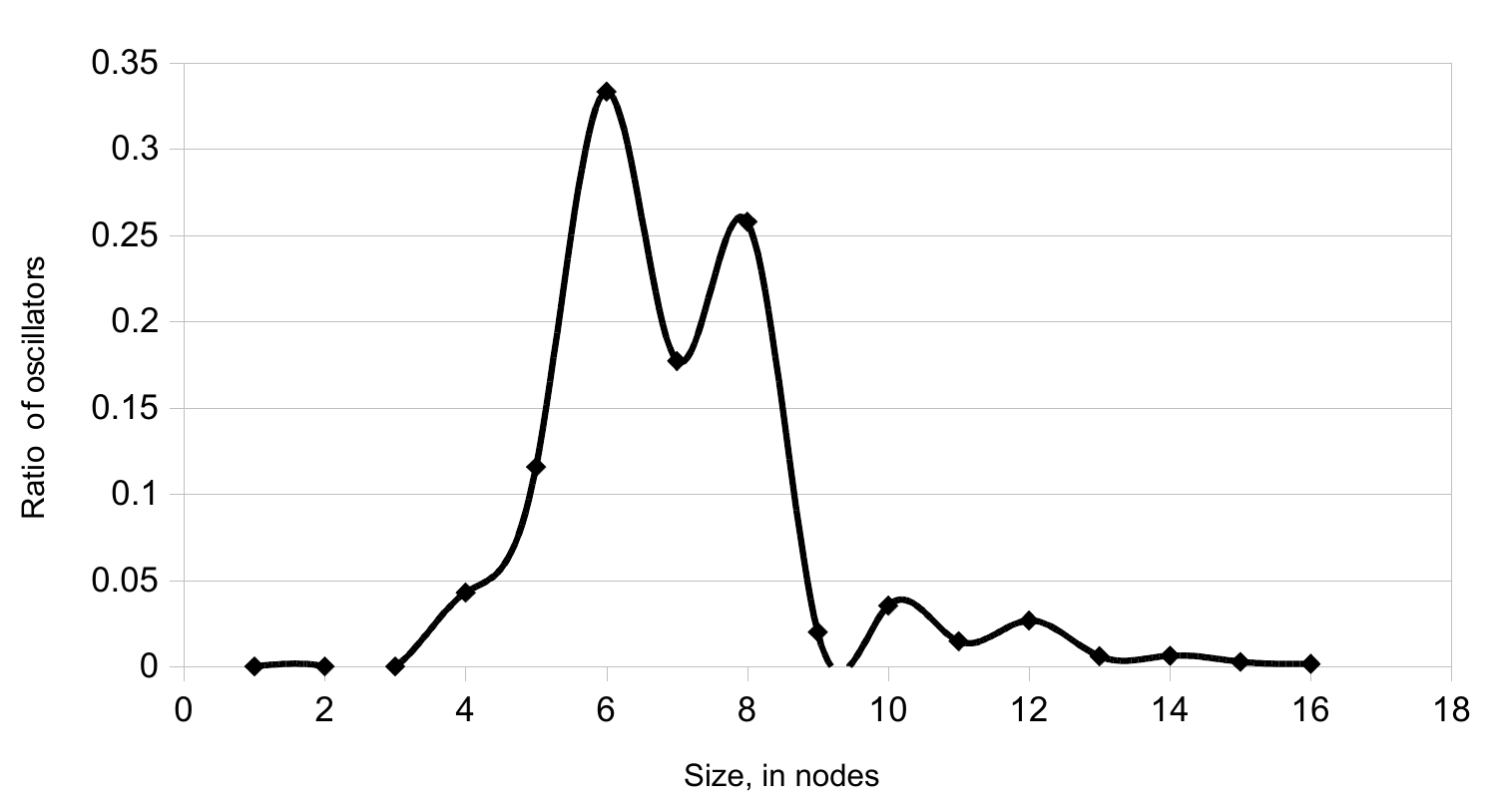}}
\subfigure[]{\includegraphics[width=0.8\textwidth]{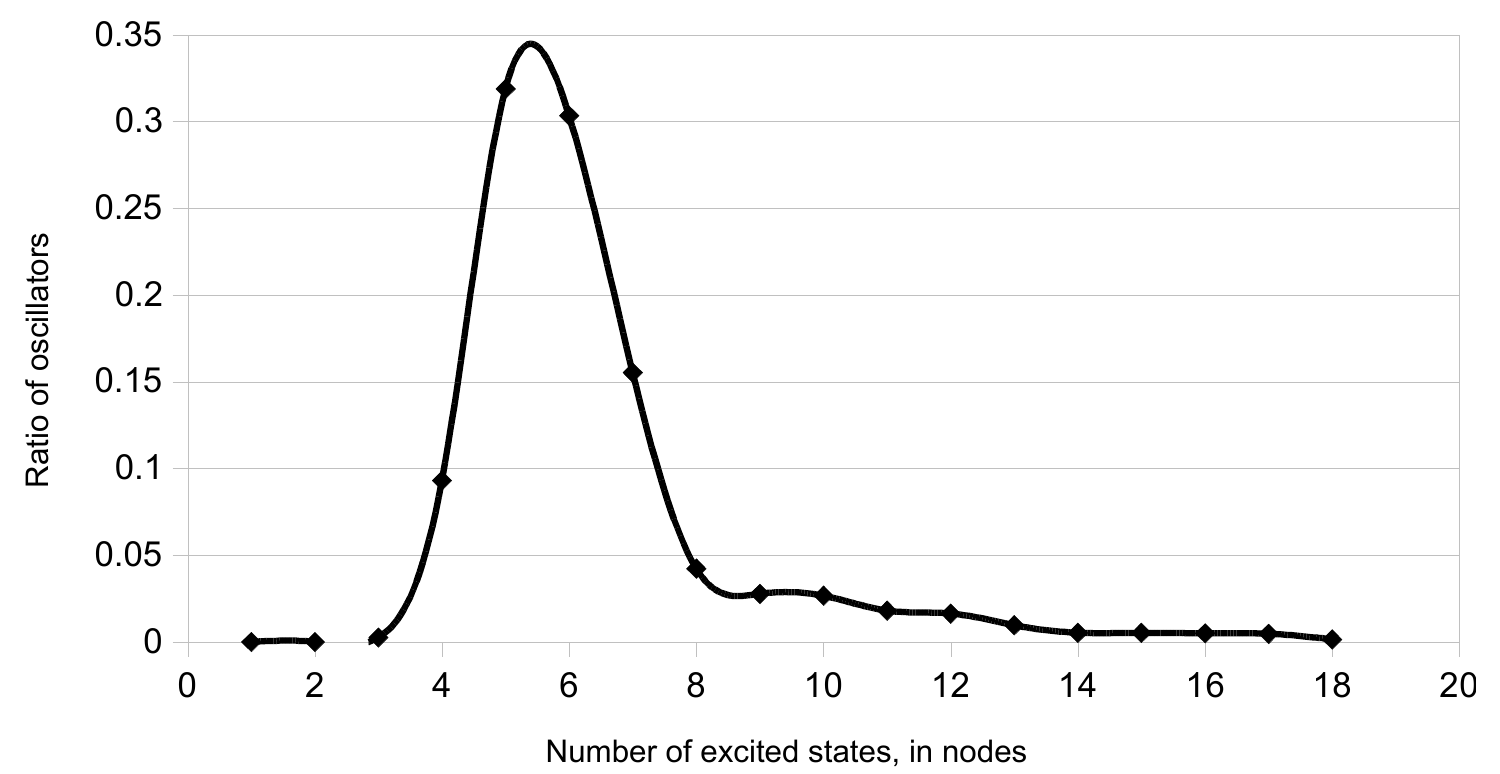}}
\caption{Average parameters of oscillators. 
(a)~Distribution of maximum linear sizes. 
(b)~Distribution of numbers of excited states in oscillators. 
The distributions were calculated on 1677 oscillating patterns 
(not necessarily different types of oscillators) discovered in 40000 trials.}
\label{graphs}
\end{figure}

\begin{figure}
\centering
\subfigure[]{
\includegraphics[scale=0.43]{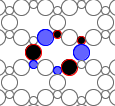}
\includegraphics[scale=0.43]{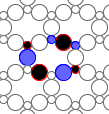}
\includegraphics[scale=0.43]{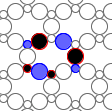}}
\subfigure[]{
\includegraphics[scale=0.43]{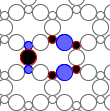}
\includegraphics[scale=0.43]{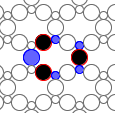}
\includegraphics[scale=0.43]{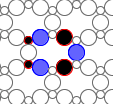}}
\subfigure[]{
\includegraphics[scale=0.43]{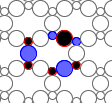}
\includegraphics[scale=0.43]{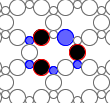}
\includegraphics[scale=0.43]{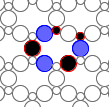}}
\subfigure[]{
\includegraphics[scale=0.43]{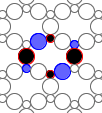}
\includegraphics[scale=0.43]{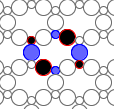}
\includegraphics[scale=0.43]{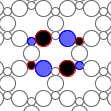}}
\subfigure[]{
\includegraphics[scale=0.43]{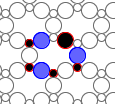}
\includegraphics[scale=0.43]{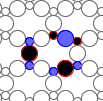}
\includegraphics[scale=0.43]{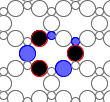}}
\subfigure[]{
\includegraphics[scale=0.44]{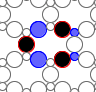}
\includegraphics[scale=0.44]{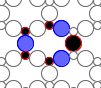}
\includegraphics[scale=0.44]{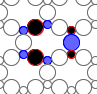}}
\subfigure[]{
\includegraphics[scale=0.43]{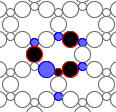}
\includegraphics[scale=0.43]{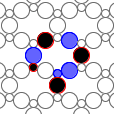}
\includegraphics[scale=0.43]{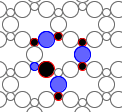}}
\subfigure[]{
\includegraphics[scale=0.43]{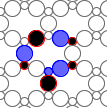}
\includegraphics[scale=0.43]{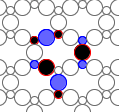}
\includegraphics[scale=0.43]{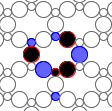}}
\subfigure[]{
\includegraphics[scale=0.43]{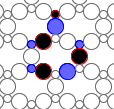}
\includegraphics[scale=0.43]{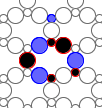}
\includegraphics[scale=0.43]{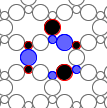}}
\subfigure[]{
\includegraphics[scale=0.43]{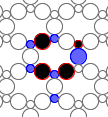}
\includegraphics[scale=0.43]{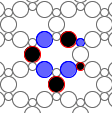}
\includegraphics[scale=0.43]{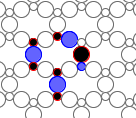}}
\subfigure[]{
\includegraphics[scale=0.43]{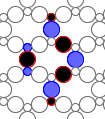}
\includegraphics[scale=0.43]{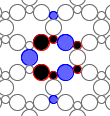}
\includegraphics[scale=0.43]{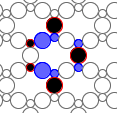}}
\subfigure[]{
\includegraphics[scale=0.43]{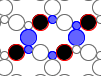}
\includegraphics[scale=0.43]{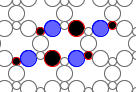}
\includegraphics[scale=0.43]{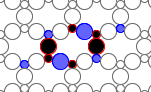}}
\subfigure[]{
\includegraphics[scale=0.43]{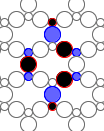}
\includegraphics[scale=0.43]{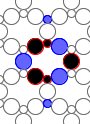}
\includegraphics[scale=0.43]{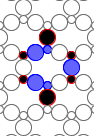}}
\subfigure[]{
\includegraphics[scale=0.43]{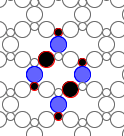}
\includegraphics[scale=0.43]{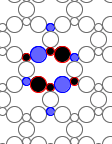}
\includegraphics[scale=0.43]{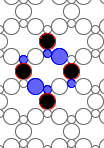}}
\caption{Exemplar configurations of oscillators with a total, i.e. over all configurations of an oscillator, 
number of excited states 12~(a--f), 14~(g--j) and 16~(k--n). All oscillators shown have 
period three, their configuration change from left to right. Excited state is black with red contour, refractory state is grey/blue.}
\label{oscillators}
\end{figure}

For excitation functions  $\sigma^t(s)>0$ and  $\sigma^t(o)>1$ ($f_{01}$),  
 $\sigma^t(s)>0$ and  $\sigma^t(o)=2$ ($f_{02}$), 
 $\sigma^t(s)=1$ and  $\sigma^t(o)>1$ ($f_{11}$),  
  and $\sigma^t(s)=1$ and  $\sigma^t(o)=2$  ($f_{12}$),
  a randomly excited automaton $\mathcal A$ evolves towards a configuration with few localised excitations.

  In terms of cellular automata, these excitations are oscillators ---  compact configurations of non-resting states undergoing modifications yet returning  to their original states in a finite number of steps (Tab.~\ref{tabpatterns}). 
  

To discover how often oscillators appear in randomly excited automaton we stimulated 
$\mathcal A$ with a rectangular domain of 10 by 10 nodes. At the beginning of evolution each node in the domain 
is assigned one of three states equiprobably. In each trial, the automaton was iterated until all excitation disappeared or 
a localised excitation emerged. For each function $f_{01}$, $f_{02}$, $f_{11}$ and $f_{12}$ 10K trials were conducted. We found that oscillators appear with probability  0.057 in automata governed by function $f_{01}$,
0.086 in automata governed by $f_{11}$, 0.0144 for $f_{02}$,  and 0.0108 for $f_{12}$.

Distributions of oscillators on linear sizes and weights (numbers of excited states) are shown in Fig.~\ref{graphs}. The distributions are obtained by averaging the maximal linear size and the number of excited states over all the 
configurations of a period of the oscillator. A typical oscillator spans five to eight nodes along one dimension and consists of four to seven excited states. Examples of minimal size and weight oscillators are shown in Fig.~\ref{oscillators}.  All oscillators discovered exhibit the same behaviour for all oscillator functions.  All oscillators studied have period three. Based on a total number $e$ of excited states, summed over three configurations of an oscillator, and a sequence of excited states 
$w=(w^t w^{t+1} w^{t+2})$ , where $w^t$ is a number of excited nodes at step $t$, in oscillator cycle, we can select the following classes of minimal oscillators:
\begin{itemize}
\item Class $C_{e=12}$. These are lightest oscillators, see examples in Fig.~\ref{oscillators}.  There are three sub-classes determined by $w$: $w^1_{12}=(444)$ (Fig.~\ref{oscillators}ad), $w^2_{12}=(345)$  (Fig.~\ref{oscillators}bc)
and $w^3_{12}=(354)$  (Fig.~\ref{oscillators}ef)
\item Class $C_{e=14}$. These are middle weight oscillators. There are three sub-classes: $w^1_{14}=(446)$  
(Fig.~\ref{oscillators}gj), and $w^2_{14}=(455)$   (Fig.~\ref{oscillators}hi).
\item Class $C_{e=16}$. These are heavy weight oscillators. There are two sub-classes:
$w^1_{16}=(466)$  (Fig.~\ref{oscillators}ln) and $w^2_{16}=(556)$  (Fig.~\ref{oscillators}km).
\end{itemize}

Ratio of excited silicon automata to excited oxygen automata in oscillators configurations are 0:3, 2:2 and 4:1 in class 
$C_{e=12}$; 1:3, 3:3 and 5:1 of class $C_{e=14}$; and, 0:4, 2:3 and 4;4 in class $C_{e=16}$. 
In class $C_{e=12}$ excitation is equally distributed between silicon and oxygen automata, due to 
equal chances of oscillator configurations with ratios 0:3 and 4:1 to appear in a randomly perturbed automaton.  
Silicon automata are excited more often in oscillators of class $C_{e=14}$, assuming equiprobability of the oscillator generation. Oxygen automata are more likely to be excited in oscillators of class $C_{e=16}$.

\begin{figure}
\centering
\subfigure[$h_1$]{
\includegraphics[scale=0.42]{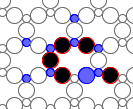}
\includegraphics[scale=0.42]{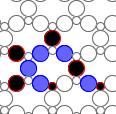}
\includegraphics[scale=0.42]{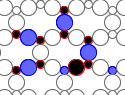}
}
\subfigure[$h_2$]{
\includegraphics[scale=0.45]{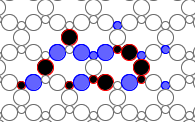}
\includegraphics[scale=0.45]{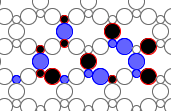}
\includegraphics[scale=0.45]{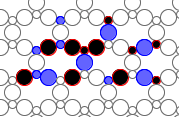}
}
\subfigure[$h_3$]{
\includegraphics[scale=0.45]{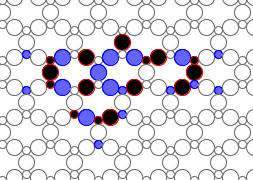}
\includegraphics[scale=0.45]{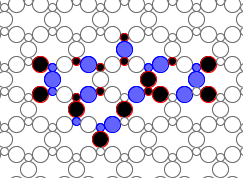}
\includegraphics[scale=0.45]{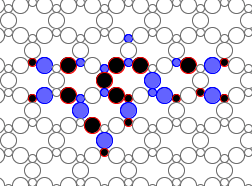}
}
\caption{Examples of heavy oscillators. Excited state is black with red contour, refractory state is grey/blue. Each oscillator has three configurations, shown in their succession from left to right.}
\label{f02oscillators}
\end{figure}

We have analysed in details only oscillating localizations with minimal size and number of non-resting node-states. 
In principle, the only limit on oscillator size is a size of automaton network. Examples of large oscillators are shown 
in Fig.~\ref{f02oscillators}. Oscillator $h_1$ change its maximum linear size between eight and nine, and consists of 
five, six and nine nodes in each of its three configurations  (Fig.~\ref{f02oscillators}a). Silicon and oxygen nodes are excited in total equal ratio.  Oscillator $h_2$ spans over 13 nodes in its largest configuration. Each configuration of $h_2$ consists of five excited silicon nodes and five or six excited oxygen nodes  (Fig.~\ref{f02oscillators}b). Oscillator $h_3$ is the largest and heaviest amongst examples in Fig.~\ref{f02oscillators}: its maximum size is 17 nodes and it has 15 and 17 excited sites in its configurations. Oxygen automata are slightly more often excited than silicon automata in patterns of $h_3$; exact ratios of $s$ to $o$ are 6:9, 8:9 and 9:8. 

\begin{figure}[!tbp]
\centering
\subfigure[$t=1$]{\includegraphics[width=0.20\textwidth]{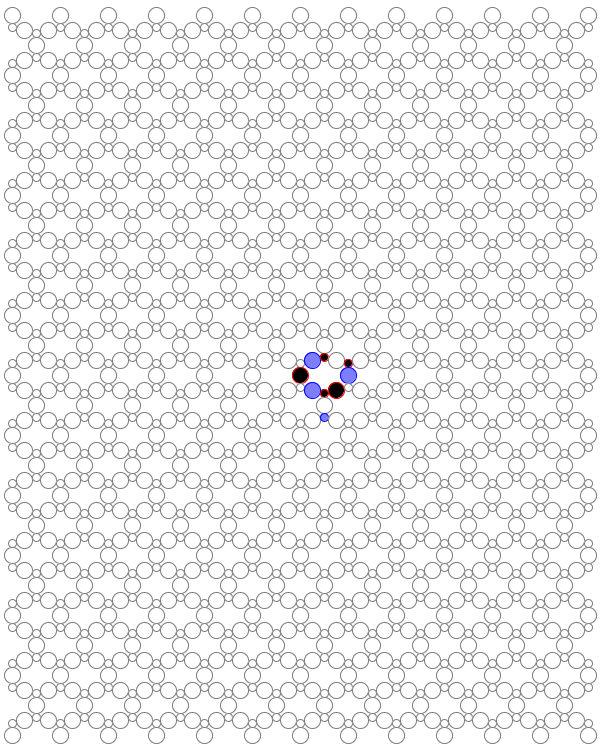}}
\subfigure[$t=2$]{\includegraphics[width=0.20\textwidth]{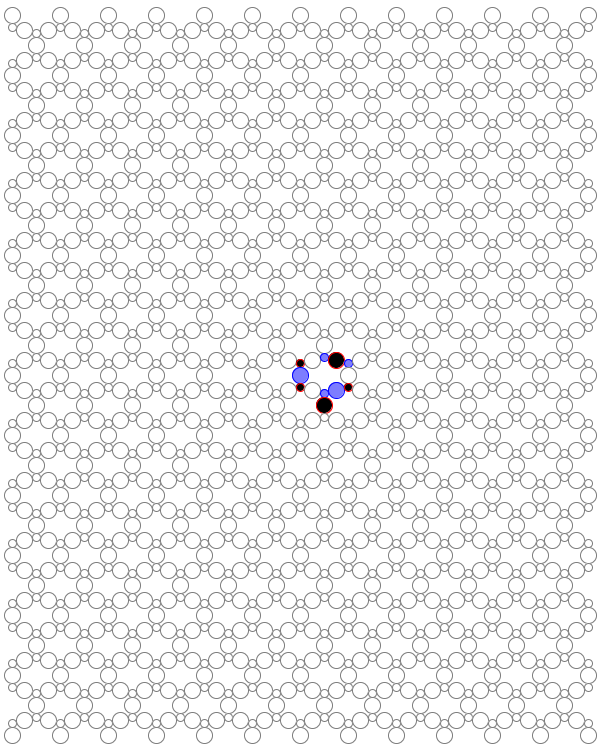}}
\subfigure[$t=3$]{\includegraphics[width=0.20\textwidth]{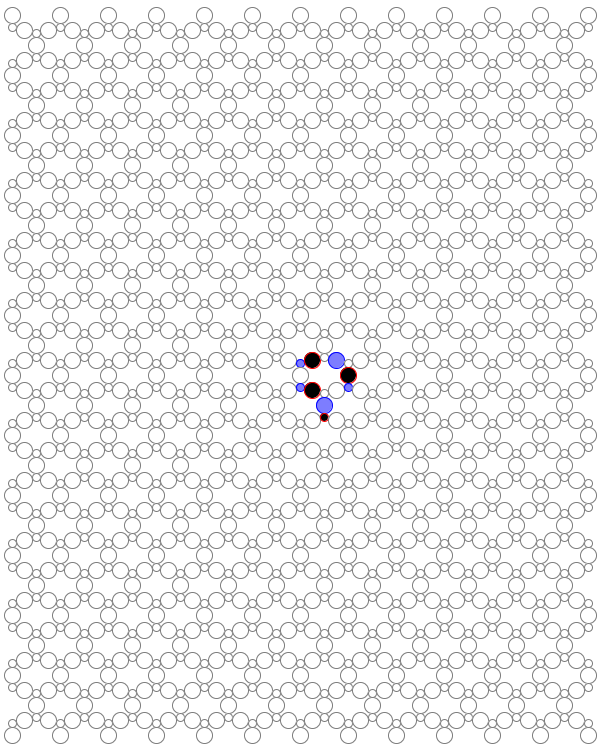}}
\subfigure[$t=4$]{\includegraphics[width=0.20\textwidth]{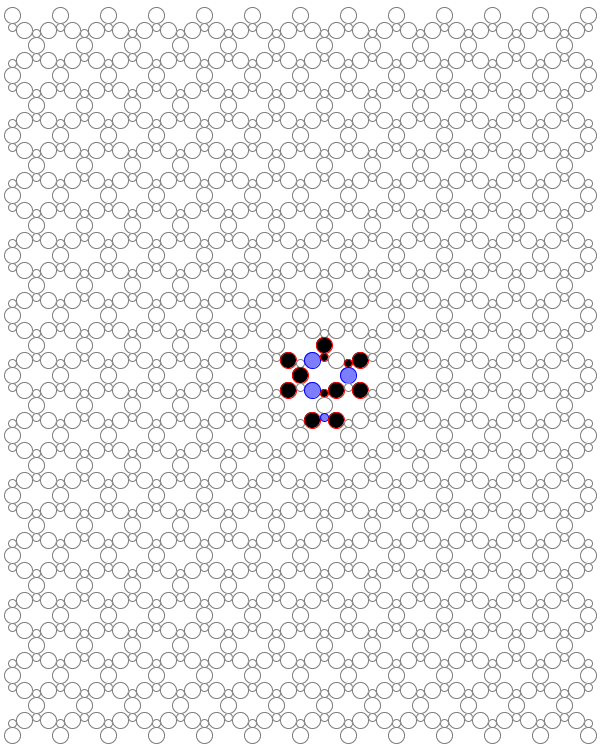}}
\subfigure[$t=5$]{\includegraphics[width=0.20\textwidth]{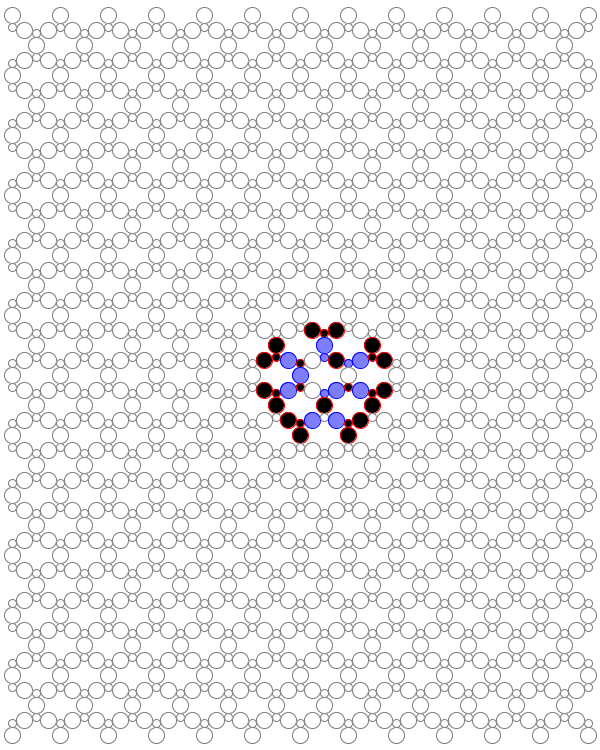}}
\subfigure[$t=8$]{\includegraphics[width=0.20\textwidth]{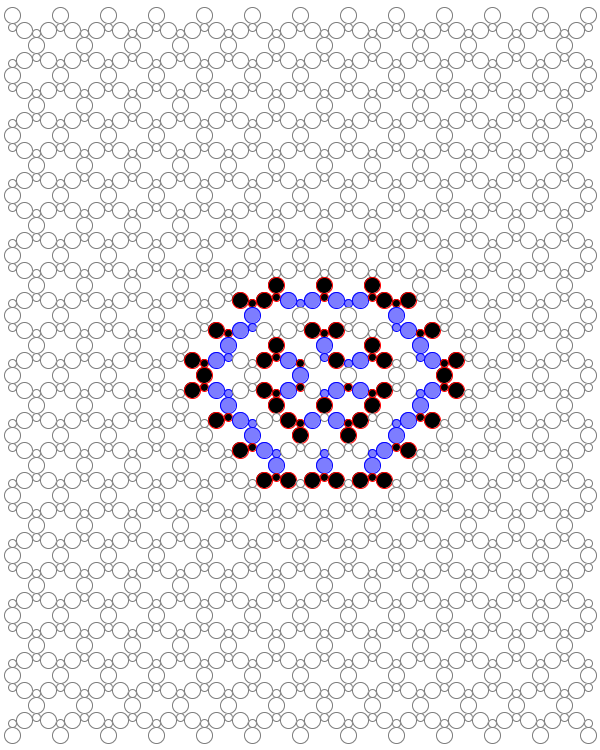}}
\subfigure[$t=9$]{\includegraphics[width=0.20\textwidth]{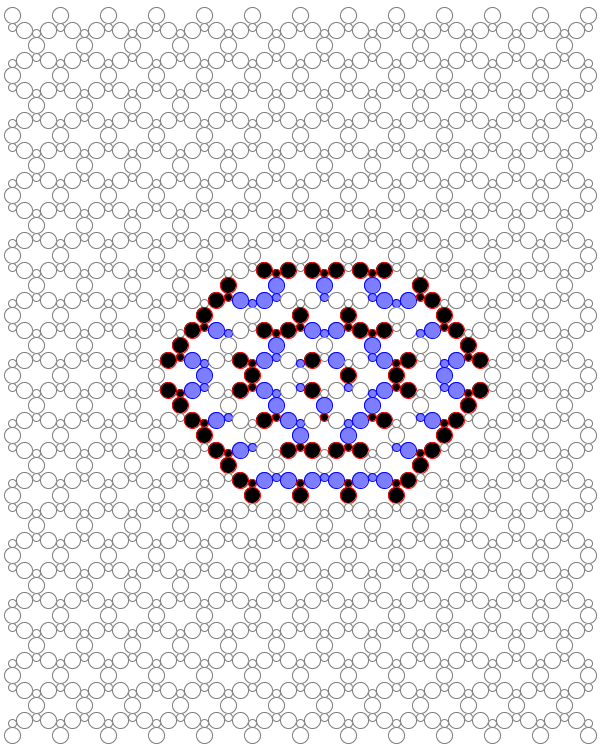}}
\subfigure[$t=10$]{\includegraphics[width=0.20\textwidth]{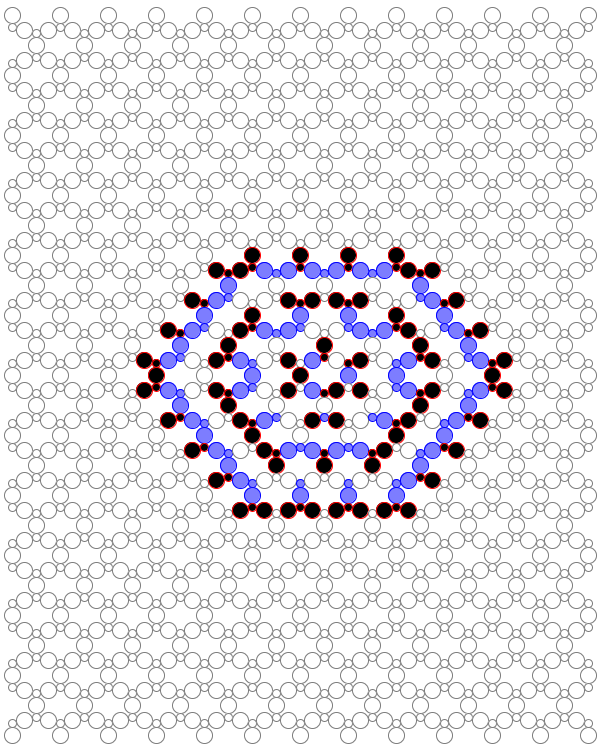}}
\subfigure[$t=16$]{\includegraphics[width=0.20\textwidth]{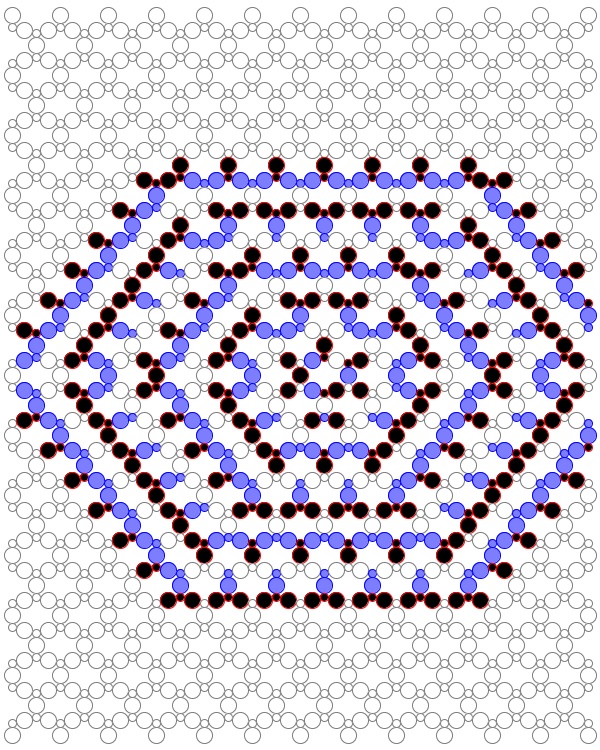}}
\subfigure[$t=19$]{\includegraphics[width=0.20\textwidth]{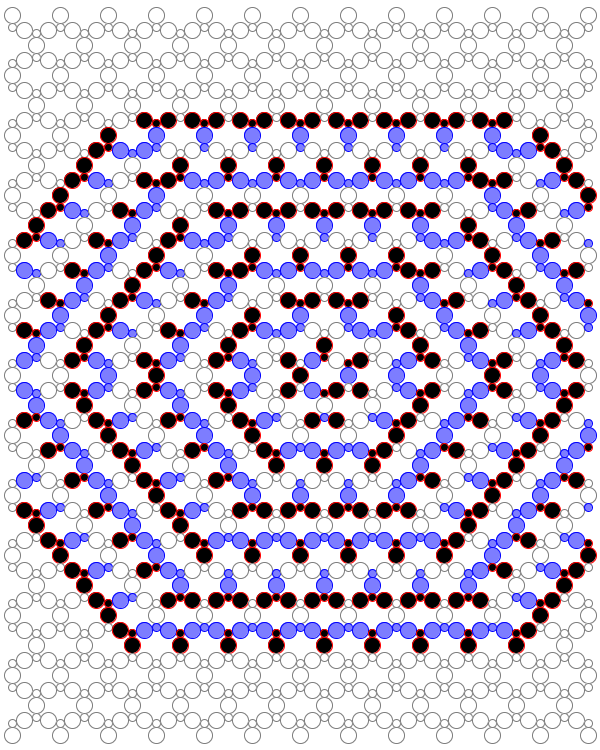}}
\subfigure[$t=20$]{\includegraphics[width=0.20\textwidth]{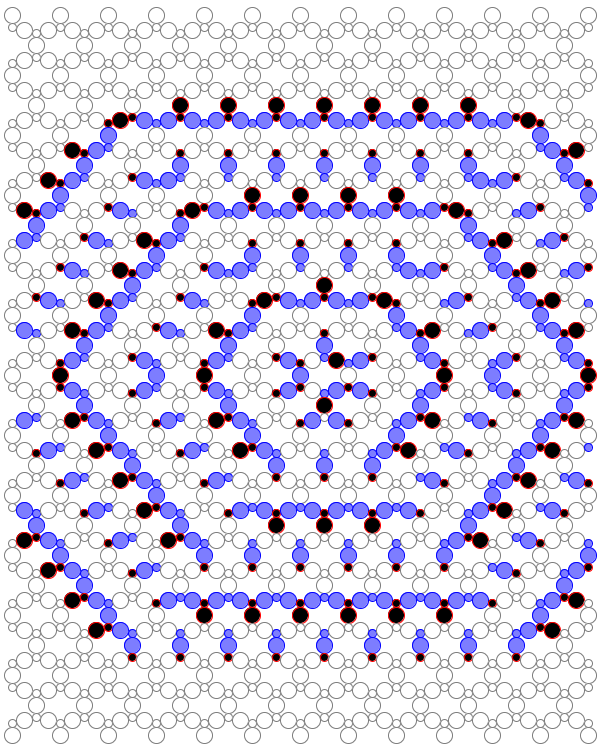}}
\subfigure[$t=21$]{\includegraphics[width=0.20\textwidth]{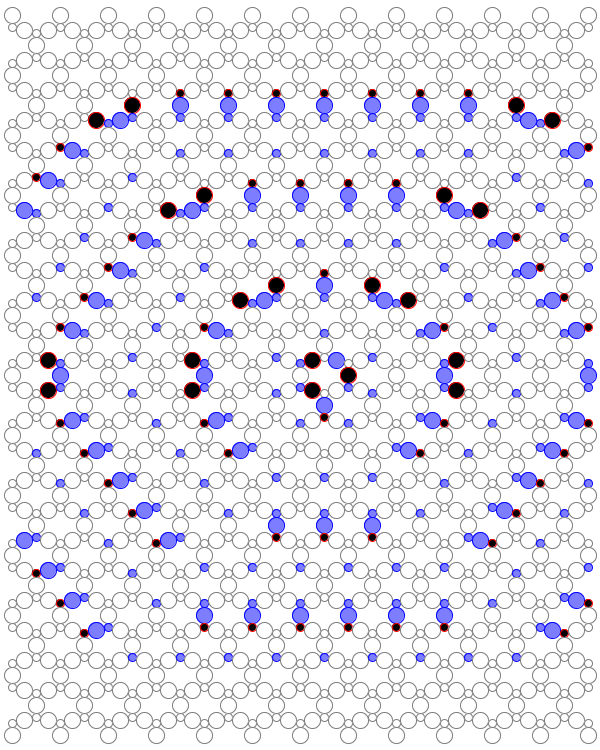}}
\subfigure[$t=22$]{\includegraphics[width=0.20\textwidth]{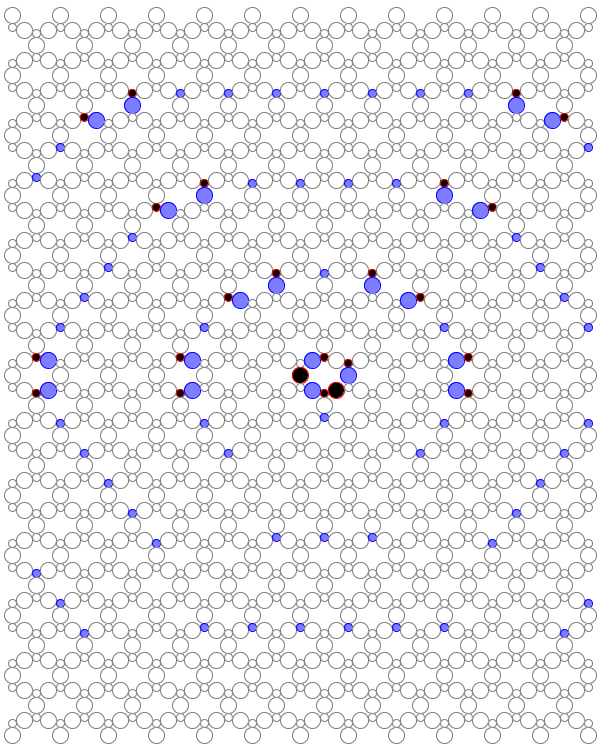}}
\subfigure[$t=23$]{\includegraphics[width=0.20\textwidth]{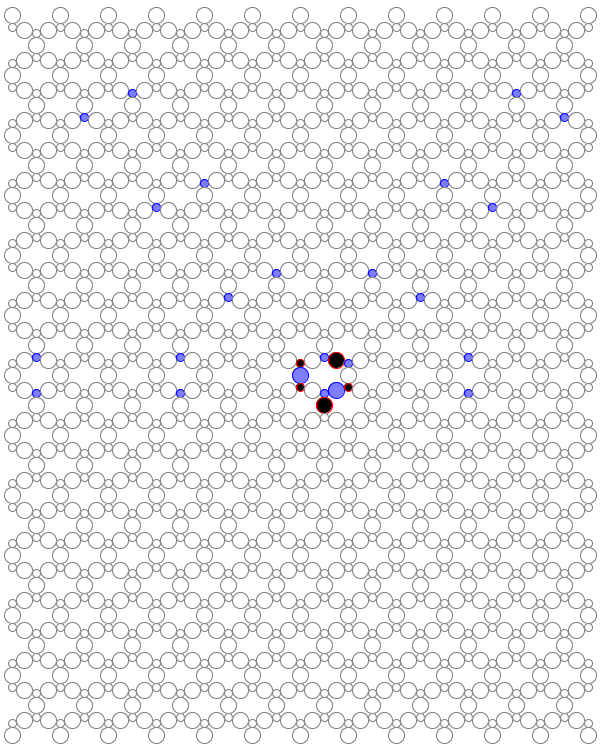}}
\subfigure[$t=24$]{\includegraphics[width=0.20\textwidth]{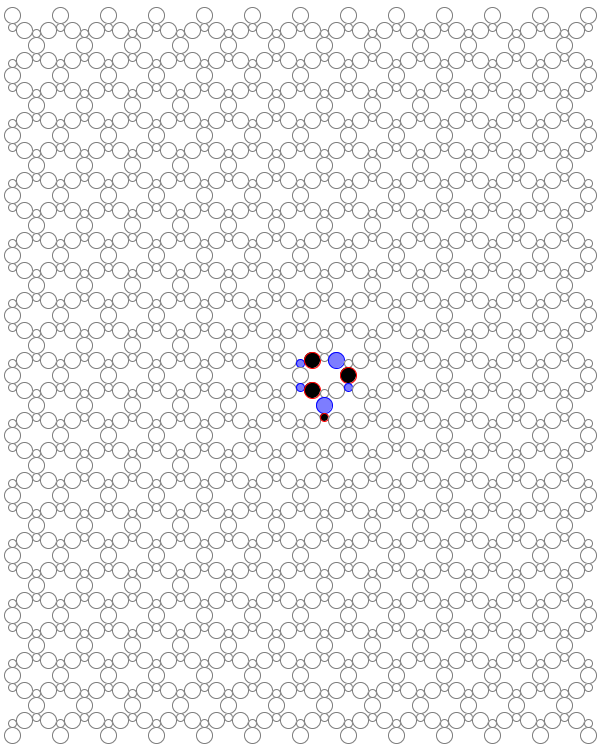}}
\subfigure[$t=25$]{\includegraphics[width=0.20\textwidth]{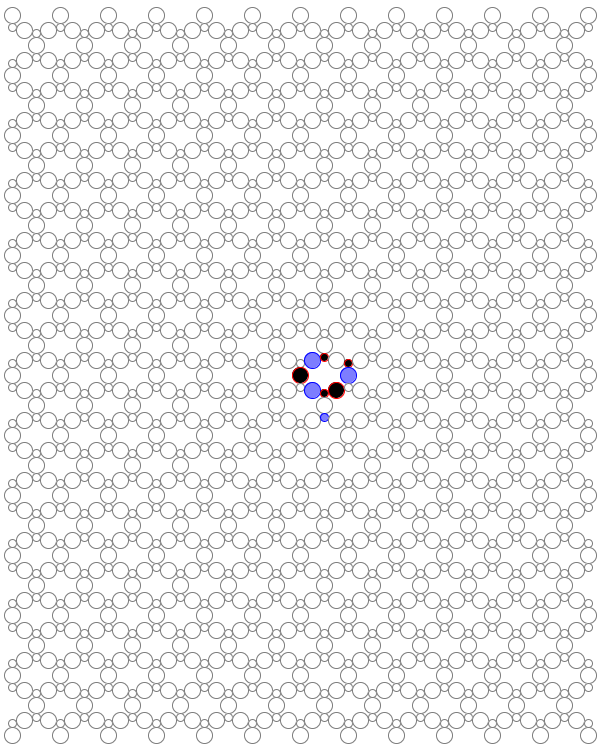}}
\caption{Dynamics of automaton $\mathcal A$, where excitation rules are changed twice during the automaton's development. Automaton evolves by rule $f_{01}$ for $t=1, \ldots, 4$.  At step t$t=4$ excitation condition is 
modified as $\sigma^t(o)>0$ (to make the wave supporting rule).  At step $t=20$ the rule is changed back to 
oscillator rule $f_{01}$. Excited states are black with red contour, refractory states are grey/blue. }
\label{osc2wave}
\end{figure}

Oscillators survive change of excitation rule from an oscillator supporting rule to a wave supporting rule and back to 
an oscillator supporting rules. When a node state transition rule is dynamically changed to a wave supporting rule an 
oscillators becomes a source of waves generation (Fig.~\ref{osc2wave}d). Resting nodes adjacent to excited sites of 
an oscillator becomes excited and excite their neighbours at the next time step. Thus waves of excitation start to emerge
around the original oscillators. The wave propagate outwards  (Fig.~\ref{osc2wave}e--n). When the node state transition rule is switched back to the oscillator supporting rule  (Fig.~\ref{osc2wave}o) the excitation waves extinguish (Fig.~\ref{osc2wave}o--q) yet original oscillator remains active (Fig.~\ref{osc2wave}r--t).

\section{Discussion}

We modelled a network of excitable finite state machines based on a simplified topological structure of 
silicate tetrahedra sheets. We studied a domain of excitation rules that support oscillating localised
excitations, provided basic classification and analysed structure of minimal oscillators. The following topics 
remain open.

All oscillators discovered so far have period three steps. Are there oscillators with period more than three? 

We did not find gliders, travelling localizations, which would propagate for a long distance without splitting into new 
localizations. Is it because the rule   (\ref{rule}) does not support gliders at all or we were unlucky in not discovering a glider in 100K random perturbations of phyllosilicate automaton?

\end{document}